\documentclass[twocolumn,showpacs,preprintnumbers,amsmath,amssymb]{revtex4-1}

\usepackage{graphicx}
\usepackage{dcolumn}
\usepackage{bm}
\usepackage{array}
\usepackage{tabularx}
\usepackage{color}
\newcolumntype{L}[1]{>{\raggedright\arraybackslash}p{#1}}
\newcolumntype{C}[1]{>{\centering\arraybackslash}p{#1}}
\newcolumntype{R}[1]{>{\raggedleft\arraybackslash}p{#1}}

\begin{document}

\preprint{311 anticrossing paper varxiv}

\title{Detection and control of spin-orbit interactions in a GaAs hole quantum point contact}

\author{A. Srinivasan$^{1}$, D. S. Miserev$^{1}$, K. L. Hudson$^{1}$, O. Klochan$^{1}$, K. Muraki$^{2}$, Y. Hirayama$^{3}$, D. Reuter$^{4}$, A. D. Wieck$^{5}$, O. P. Sushkov$^{1}$, and A. R. Hamilton$^{1}$}
\email{Alex.Hamilton@unsw.edu.au}
\affiliation{$^{1}$School of Physics, University of New South Wales, Sydney NSW 2052, Australia}
\affiliation{$^{2}$NTT Basic Research Laboratories, NTT corporation, Atsugi-shi, Kanagawa 243-0198, Japan}
\affiliation{$^{3}$Graduate School of Science, Tohoku University, Sendai-shi, Miyagi 980-8578 Japan}
\affiliation{$^{4}$Fachbereich Physik, University of Paderborn, Warburger Stra{\ss}e 100, 33098 Paderborn, Germany}
\affiliation{$^{5}$Angewandte Festkorperphysik, Ruhr-Universit{\"a}t Bochum, D-44780 Bochum, Germany}

\date{\today}

\begin{abstract}
We investigate the relationship between the Zeeman interaction and the inversion asymmetry induced spin orbit interactions (Rashba and Dresselhaus SOIs) in GaAs hole quantum point contacts.  The presence of a strong SOI results in crossing and anti-crossing of adjacent spin-split hole subbands in a magnetic field.  We demonstrate theoretically and experimentally that the anti-crossing energy gap depends on the interplay between the SOI terms and the highly anisotropic hole $g$ tensor,  and that this interplay can be tuned by selecting the crystal axis along which the current and magnetic field are aligned.  Our results constitute independent detection and control of the Dresselhaus and Rashba SOIs in hole systems, which could be of importance for spintronics and quantum information applications.
\end{abstract}

\maketitle

The spin-orbit interaction (SOI) couples the spin of a charged particle to its orbital motion, opening up the possibility of using electric fields to manipulate spin-related properties.  This concept is central to the emerging field of spintronics where the ultimate goal is the development of low dissipation spin-based transistors and spin-orbit qubits for quantum computation \cite{DattaApl90, LossPRA98, WolfSci01, Awsbook02}.

Hole systems in GaAs offer an attractive platform for electrical spin control, due to the non-zero orbital angular momentum in the valence band, where the heavy hole states possess spin 3/2 \cite{WinklerBook03}.  When heavy holes are quantum confined to a two-dimensional plane, there is a strong spin-orbit interaction arising from inversion asymmetry in both the crystal and the confinement potential - the Dresselhaus SOI \cite{Dress} and Rashba SOI \cite{Rashba, WieckPRL} respectively.  Quantifying and controlling these interactions is of interest for hole spin qubits \cite{DlossPRL07, GoloPRB06}, while appropriate tuning of the Rashba and Dresselhaus terms allows the formation of long-lived persistent spin-helix states \cite{Helix, 111}.  
However, although it has been possible to disentangle the effects of the Rashba and Dresselhaus interactions in electron systems \cite{Sasaki}, there has yet to be any demonstration of independent measurement and control of these interactions in hole systems despite a number of theoretical proposals \cite{DlossPRB05, AwongPRB10}.


In this work, we study the Rashba and Dresselhaus interactions in a hole system.  We are able to separate the influence of the two interactions by comparing their effect on the Zeeman spin splitting of 1D hole subbands in GaAs quantum point contacts on different crystal planes.  While the Rashba SOI is approximately independent of the crystal plane, the Dresselhaus SOI has a strong crystal dependence. The strength and nature of the Rashba and Dresselhaus interactions controls the magnitude of the anti-crossings as adjacent 1D hole levels approach each other when an in-plane magnetic field is applied. We find that the Rashba SOI is much stronger than the Dresselhaus SOI on (100) substrates, but the two are comparable on (311) substrates. Additionally, we show that it is possible to control the strength of the Dresselhaus SOI by changing the QPC quantisation axis (current direction).  Our results are consistent with a theoretical model that incorporates crystal anisotropies, where the anti-crossing gap arises as a result of the Dresselhaus and Rashba SOIs and the tensor structure of the anisotropic hole $g$ factor.

The devices used in this work are fabricated from two GaAs/AlGaAs accumulation mode heterostructures \cite{ClarkeJAP06}, one of which is grown on the high symmetry (100) plane, and the other on the low symmetry (311)A plane. The 2D hole system (2DHS) on both heterostructures is formed in an approximately triangular confinement potential, where the large electric field results in a strong Rashba SOI. The 2DHS, which has a density $p \sim 1.5 \times 10^{11} cm^{-2}$, is further confined using 400 $\times$ 400nm split-gates to form a quantum point contact (QPC).  The (311) device has two orthogonal QPCs along the [$\overline{2}33$] and [$01\overline{1}$] crystal directions, which we label QPC[$\overline{2}33$] and QPC[$01\overline{1}$], as shown in Fig.1.  The (100) device also has two orthogonal QPCs, oriented along [$011$] and [$01\overline{1}$]. Due to the high symmetry of the (100) plane both of these QPCs show identical results  \cite{ChenNJP10}, so only the QPC oriented along $[011]$ (labelled QPC[011]) is used for this work.

Measurements were carried out in a dilution refrigerator, with a base temperature below 40mK, using standard ac lock-in techniques.  Conductance traces for all three QPCs, exhibiting characteristic 1D quantised plateaus as a function of side gate voltage $V_{SG}$, are shown in Fig.1 a-c.  The first subband shows anomalous features which may be due to the 0.7 structure \cite{MicoJPCM11, OlgaPRL14, IqbalNat13}, which will be studied in detail elsewhere.  


\begin{figure}
\includegraphics[width=0.99\linewidth]{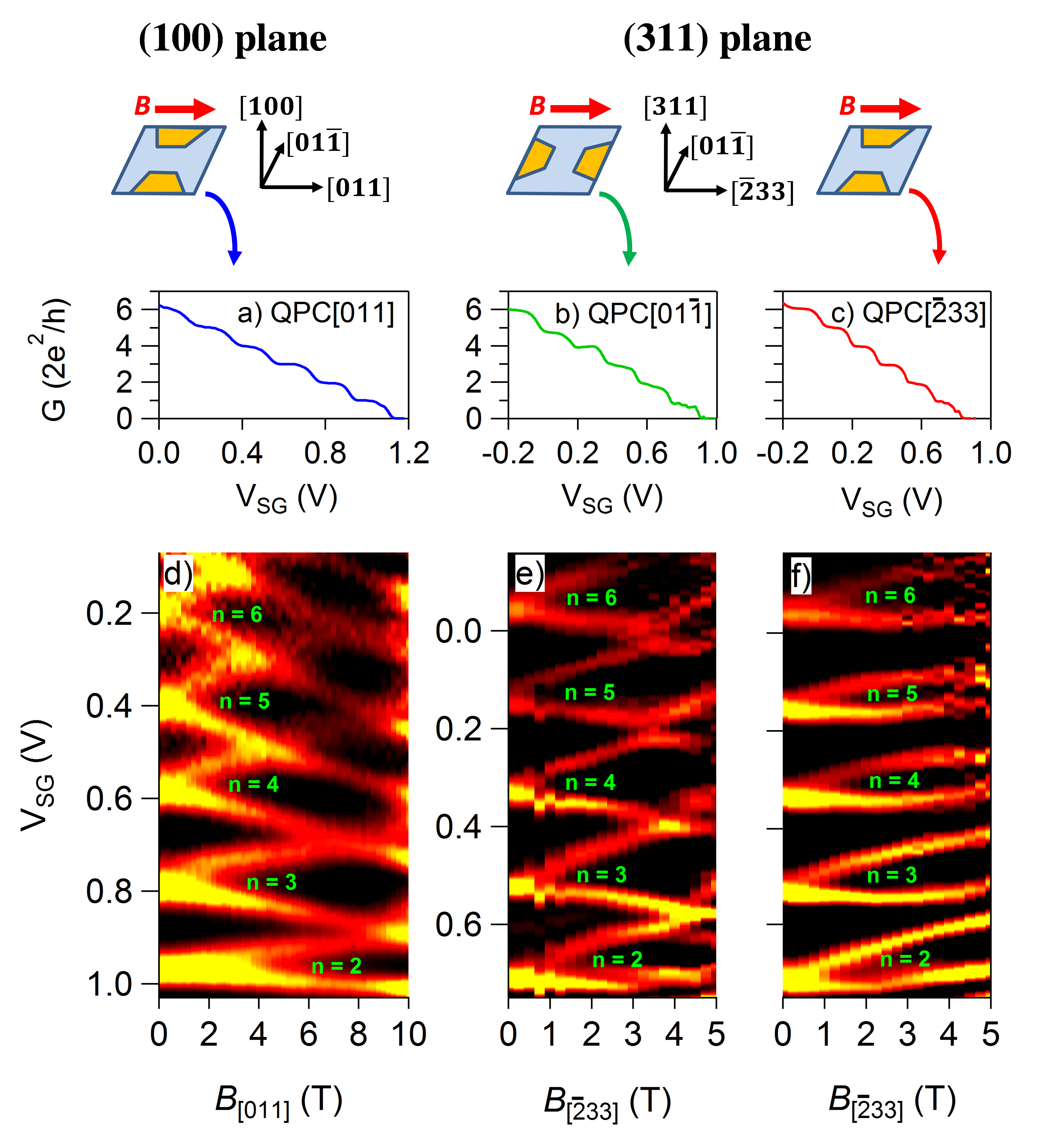}
\caption{The top panel shows schematic diagrams of the crystal orientation of the QPCs and the applied magnetic field.  (a-c) Conductance G versus $V_{SG}$ for the three QPCs, showing characteristic 1D conductance plateaus. (d-f) Zeeman splitting of 1D subbands in an in-plane magnetic field for all three QPCs. Color plots of the transconductance $\partial G/\partial V_{SG}$ are shown, where the light regions represent the 1D subband edges.  d) shows results for the (100) device where there is a linear Zeeman splitting with clear crossings between adjacent 1D subbands.  e) and f) show results for the (311) device: in e), the QPC is parallel to $[01\overline{1}]$, and anti-crossing between adjacent 1D subbands is present but weak. In f), the QPC is parallel to $[\overline{2}33]$ and strong anti-crossings are observed.} 
\end{figure}

Zeeman splitting of 1D hole subbands in an in-plane field was measured for all three QPCs, and is presented in Fig.1 (d-f).  The experimental data is presented as color plots of the transconductance $\partial G/\partial V_{SG}$, where light regions represent high transconductance, corresponding to the 1D subband edges.  Fig.1d depicts the simplest case of QPC[011] on the (100) plane, with the magnetic field applied parallel to the current, along [$011$].  The 1D subbands show linear spin splitting at low fields, and then exhibit crossings at higher fields when the spin splitting becomes larger than the 1D subband spacing \cite{ChenNJP10, AshwinNL13, KomijaniEPL13, NichelePRL14}.  These crossings indicate that there is no coupling between adjacent 1D subbands.

Now we turn to the more interesting case of the (311) device:  Here, the magnetic field was applied along $[\overline{2}33]$.  In Fig.1e, QPC$[01\overline{1}]$ also shows linear spin splitting at low fields.  However, at higher fields, adjacent subbands show weak anti-crossing behaviour (clearly visible for subbands 4-6), suggesting mixing between adjacent 1D states of opposite spin.  Finally, for QPC$[\overline{2}33]$ (Fig.1f), the anti-crossing is very strong and clearly visible between all subbands.

We note that both the 2D and 1D confinement potentials, and hence the Rashba SOI, are nearly identical for the QPCs on the (100) and (311) heterostructures.  Therefore, the different anti-crossing behaviour in Fig.1(d-f) must be due to the differing Dresselhaus SOI and Zeeman terms.  

To confirm the role of the Dresselhaus interaction, we model the system using an effective Hamiltonian approach, describing only the ground heavy hole sub-band. The influence of higher sub-bands is accounted for via perturbation theory up to the second order of the in-plane momentum $k$. 

The simple case of a QPC on the (100) plane is considered first.  Using the coordinate system $(x,y,z) = ([011], [01\overline{1}], [100])$, the effective Hamiltonian is:
\begin{eqnarray}
&& H_{[100]} = k^2/(2 m^*) + U(x,y) - \frac{B_x}{2} g_{xx} \sigma_x \nonumber \\
&& - \frac{B_y}{2} g_{yy} \sigma_y - \frac{i \alpha}{2} (k_+^3 \sigma_- - k_-^3 \sigma_+) \
\label{heff100}
\end{eqnarray}
The first term is the hole dispersion within the effective mass approximation, with $m^* = 0.2 m_e$.  The second term is the parabolic 1D confining potential of the QPC.  The third and fourth terms are the Zeeman terms for 2D holes in an in-plane magnetic field.  The fifth term is the Rashba SOI \cite{WinklerBook03}.  The Dresselhaus SOI is negligibly small in the (100) plane and is ignored here (see supplemental material \cite{suppmat} and Ref.\cite{Liz} for a description of the Rashba and Dresselhaus SOI).

For QPC[011], we define [011] as $\hat{x}$. We set the momentum along the channel $k_x$ to 0 (since we measure spin-splitting at the 1D subband edges). The Rashba term then simplifies to $- \alpha k_y^{3}\sigma_{x}$.  When the magnetic field is also applied along $[011]$, the Zeeman term $B_x g_{xx} \sigma_{x}$ is parallel to the Rashba term, so anti-crossings are prohibited \cite{NichelePRL14}.

Next we consider the case of an orthogonal QPC along the $[01\overline{1}]$ (y) direction with the field applied perpendicular to the QPC, i.e $B_x$ is applied along [011].  In this case, we set $k_y$ to 0, and the remaining odd-momentum terms in the effective Hamiltonian in eqn.~(\ref{heff100}) are considered as a small perturbation that take the following form:
\begin{equation}
V =  - \alpha k_x^3 \sigma_y \
\label{V}
\end{equation}
This perturbation (arising from the Rashba SOI) would lead to an anti-crossing of states from adjacent 1D subbands with different spin orientations.  However, for (100) QPCs, an in-plane field applied perpendicular to the QPC does not cause the 1D levels to split ($g^{*}_{\perp} \simeq 0$) \cite{ChenNJP10,KomijaniEPL13,NichelePRL14}, so anti-crossings have never been observed. 

To obtain a finite $g^{*}_{\perp}$, we now turn to the QPCs on the (311) heterostructure. 
Here the effective Hamiltonian contains an additional Zeeman term due to the off-diagonal component of the $g$ tensor, $g_{xz}$ \cite{WinklerSST08, YeohPRL14, AshwinPRB16}, and an additional Dresselhaus term, $D_1$ (see supplemental material \cite{suppmat}). Taking the coordinate system $(x,y,z) = ([\overline{2}33], [01\overline{1}], [311])$, the effective Hamiltonian is now:
\begin{eqnarray}
&& H_{[311]} = k^2/(2 m^*) + U(x,y) - \frac{B_x}{2}(g_{xz} \sigma_z + g_{xx} \sigma_x) \nonumber \\
&& - \frac{B_y}{2} g_{yy} \sigma_y - \frac{i \alpha}{2} (k_+^3 \sigma_- - k_-^3 \sigma_+) - D_1 k_y \sigma_z \
\label{heff}
\end{eqnarray}

We start with the QPC parallel to the $[01\overline{1}]$ (y) direction, and set $k_y = 0$.  As before the odd momentum terms take the form of eqn.(\ref{V}) when the magnetic field is applied perpendicular to the QPC, i.e. $B_x$ is applied along $[\overline{2}33]$.  In this case the corresponding spinors $|\pm \rangle$ should be eigenvectors of the matrix $G = g_{xz} \sigma_z + g_{xx} \sigma_x$ according to eqn.(\ref{heff}).
The anti-crossing gap between the $n^{th}$ and $n+1^{th}$ subbands is then:
\begin{equation}
\Delta(n) = 2\left|\alpha \langle k_x^3 \rangle \right| \
\label{xy}
\end{equation}
where $\langle ... \rangle = \langle n | ... | n+1 \rangle $.	
This gap (of order $\sim 100 \mu eV$ \cite{suppmat}) appears solely due to the Rashba coupling constant, since the $D_1$ Dresselhaus contribution vanishes when $k_y = 0$. 

Finally, we consider the QPC parallel to the $[\overline{2}33]$ (x) direction and set $k_x$ to 0.
In this case, the Dresselhaus SOI is preserved and the odd-momentum correction takes the form:
\begin{equation}
V =  - \alpha k_y^3\sigma_x - D_1 k_y \sigma_z \
\end{equation}
If the magnetic field $B_x$ is again applied along $[\overline{2}33]$, the anti-crossing gap now becomes:
\begin{equation}
\Delta(n) = 2\left|\alpha \langle k_y^3 \rangle \frac{ g_{xz}}{g} - D_1 \langle k_y \rangle \frac{g_{xx}}{g} \right|
\label{xx}
\end{equation}
where $g = \sqrt{g_{xx}^2 + g_{xz}^2}$ is the total $g$-factor.	
In the limit $k = 0$, $g_{xz}$ and $g_{xx}$ have different signs \cite{WinklerSST08}, and the Rashba and Dresselhaus terms in (\ref{xx}) add constructively.  Anti-crossing gaps for this case are predicted to be approximately two times greater than for QPC $\parallel [01\overline{1}]$, due to the additional $D_1$ Dresselhaus contribution, which is of similar magnitude to the Rashba term for the (311) plane \cite{suppmat}.  

We now use Eqs. 1-6 to produce a simple qualitative picture of the Zeeman spectrum of the three QPCs, presented in Fig.2.  The lines show the evolution of the 1D subbands as a function of magnetic field.  For simplicity, we neglect subband dependence of the $g$ tensor and the inter-subband spacing \cite{AshwinPRB16}, to show that the variation in anti-crossing gap between the three QPCs arises solely due to the differing coupling between the SOI and Zeeman terms.  Details of the calculations are given in the supplemental material \cite{suppmat}.  

\begin{figure}
\includegraphics[width=0.99\linewidth]{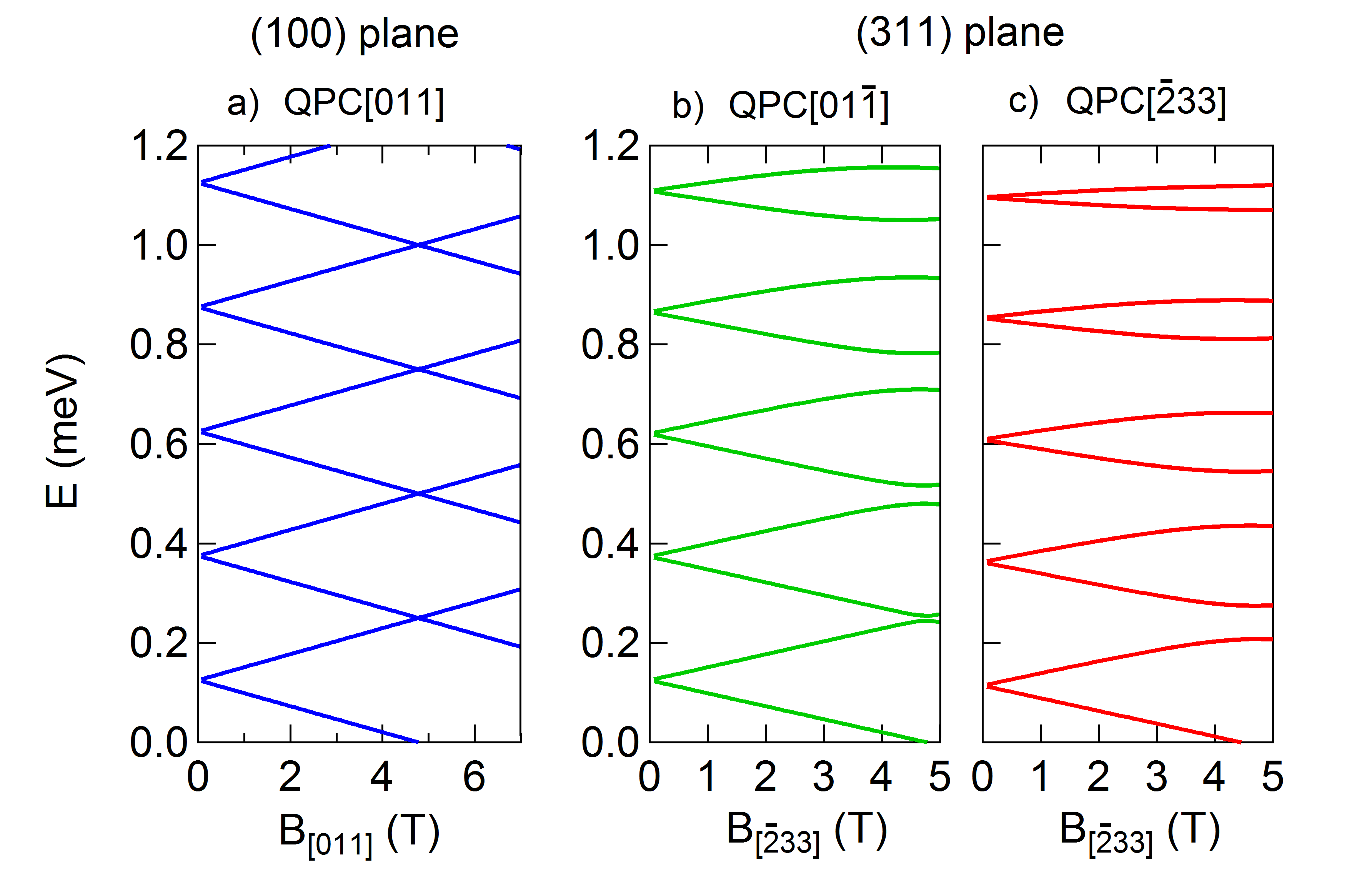}
\caption{Zeeman spectrum calculated using Eqs. 1-6 for the three QPC orientations.  The magnitude of the anti-crossing gaps for the three QPCs are in qualitative agreement with the corresponding experimental data in Fig.1(d-f). The 1D subband spacing is $0.25meV$, $g_{xz} = 0.75$, $g_{xx} = g_{yy} = -0.5$, Rashba coupling $\alpha = 400$ eV\AA$^3$, and Dresselhaus coupling $D_1 = 28 meV$\AA~\cite{suppmat}.}
\end{figure}

The calculated energy spectrum shows no anti-crossing for QPC[011] on the (100) plane (Fig.2a), since there is no interaction between adjacent subbands.  For QPC$[01\overline{1}]$ on the (311) plane (Fig.2b), small anti-crossing gaps appear due to the Rashba SOI.  Finally, for QPC$[\overline{2}33]$ (Fig.2c), large anti-crossings occur due to the combination of the Rashba and Dresselhaus SOI.
Figs.2b and 2c also predict that the anti-crossing energy gaps increase with subband index, due to the dependence on $\langle k^{3}\rangle$ in Eqs. (\ref{xy}) and (\ref{xx}) (see supplemental material \cite{suppmat}), which is qualitatively consistent with the data in Fig.1(d-f).  Overall, the theoretically predicted Zeeman spectrum is remarkably consistent with the experimental data, suggesting that the experimental values of the Rashba and Dresselhaus constants are close to the theoretical values of $\alpha \approx 400$ eV\AA$^3$, and $D_1 \approx 28 meV$\AA~used in the calculations \cite{suppmat}.


The large anti-crossing gap observed for QPC$[\overline{2}33]$ is evidence for the strong Dresselhaus SOI in (311) GaAs hole systems.  Furthermore, the results in Figs. 1e and 1f also show that the effect of the Dresselhaus SOI can be independently switched on or off in a QPC by selecting the quantisation axis (current) direction.  Control of the SOI is of interest for potential applications to hole based spin-qubits, suppression of spin relaxation and formation of spin helix states \cite{DlossPRL07, GoloPRB06, Helix, 111}. 

Finally, to complete the characterisation of both SOI terms and their interplay with the Zeeman spin splitting, we also investigate the other two magnetic field directions: The other orthogonal in-plane field direction ($B_y$) is considered in the supplemental material \cite{suppmat}, and the out-of-plane magnetic field direction ($B_z$) is discussed in the following paragraphs.

Fig.3 shows transconductance data with an out-of-plane field $B_z$, applied along [311] for QPC$[\overline{2}33]$. As expected for this field direction, the data shows very large Zeeman splitting due to the large $g_{zz}$ term and an upward curvature of the subbands due to additional magnetic confinement (see supp. info of ref. \cite{AshwinPRB16} and \cite{AshwinNL13, NichelePRL14}).  We also see evidence for anti-crossing between 1D levels shown by white arrows.  For the (311) plane, the anti-crossing in an out-of-plane field is analogous to that which occurs in an in-plane field: When the field is applied along [311] (z) instead of $[\overline{2}33]$ (x), the corresponding Zeeman term, $B_z g_{zz} \sigma_{z}$, plays the same role as the $B_x g_{xz} \sigma_{z}$ term in Eq.(\ref{heff}).  

\begin{figure}
\includegraphics[width=0.99\linewidth]{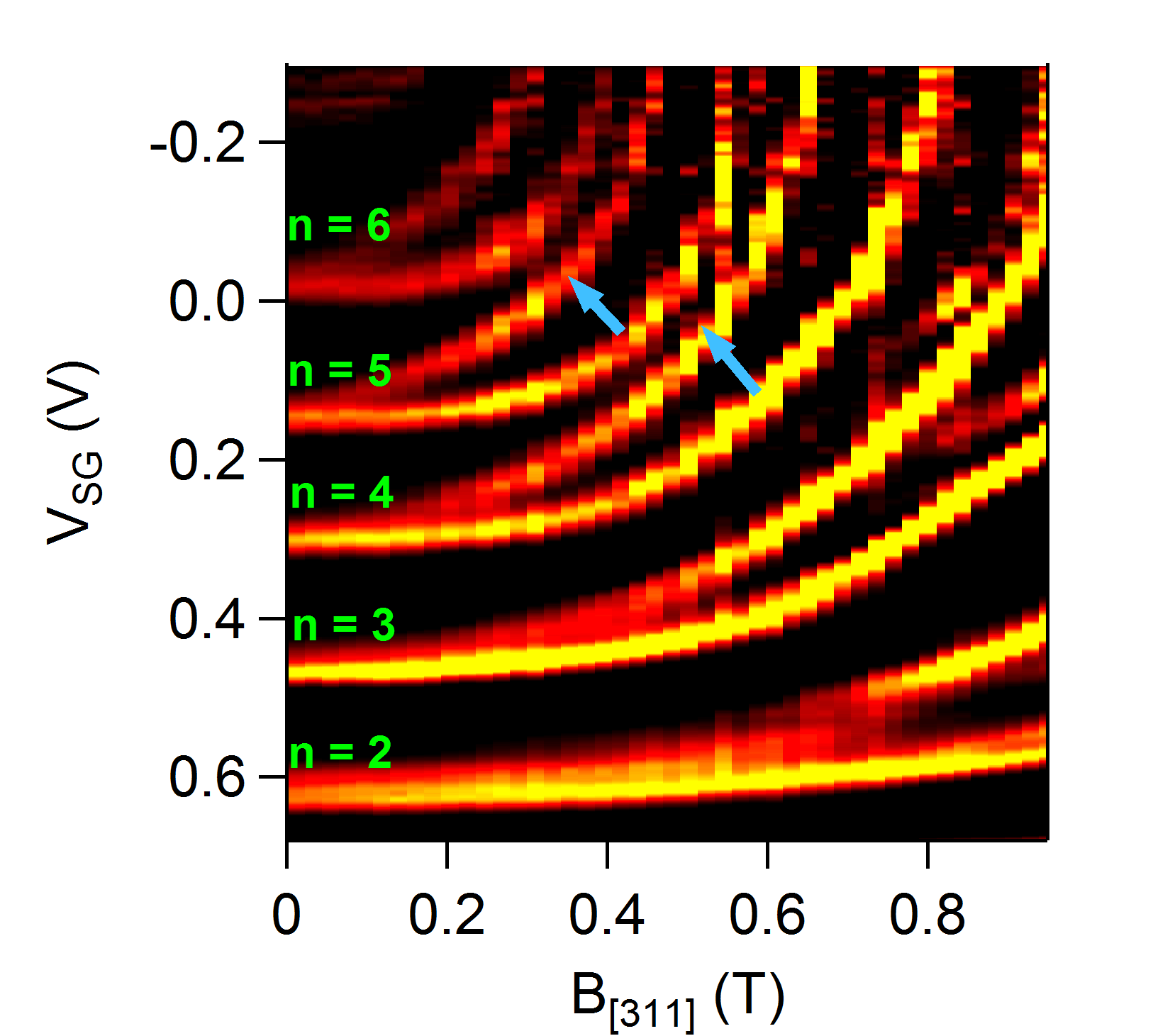}
\caption{Zeeman splitting of 1D subbands in an out-of-plane magnetic field for QPC$[\overline{2}33]$. This data also shows evidence for anti-crossing, as indicated by the blue arrows.  The subbands also exhibit a curvature to higher energies due to the additional magnetic confinement.}
\end{figure}

We can therefore very simply derive the expected anti-crossing gap for a general tilted field in (311) GaAs:
Let us consider QPC$[\overline{2}33]$ in a magnetic field applied in the xz-plane, where $(x,y,z) = ([\overline{2}33], [01\overline{1}], [311])$:
\begin{equation}
\left\{
\begin{array}{cc}
B_x = B \cos \phi\\
B_z = B \sin \phi
\end{array}
\right.
\end{equation}
Here the angle $\phi$ is assumed to be fixed.  A magnetic field along [311] (z) gives an additional Zeeman term to the effective Hamiltonian in Eq.(\ref{heff}): $-g_{zz} B_z \sigma_z/2$.  The resulting effective Hamiltonian is similar to (\ref{heff}) if we substitute:
\begin{equation}
\left\{
\begin{array}{lll}
g_{xx} & \to & g_{xx} \cos \phi\\
g_{xz} &\to & g_{xz} \cos\phi + g_{zz} \sin \phi
\end{array}
\right.
\label{subs}
\end{equation}

The corresponding anti-crossing gaps are given by Eq.~(\ref{xx}) where $g_{xx}$ and $g_{xz}$ are substituted according to~(\ref{subs}).  Note that for an out-of-plane field, the off-diagonal $g_{xz}$ term is not required to produce the anti-crossing.  Therefore, in the presence of a Rashba SOI, anti-crossing in an out-of-plane magnetic field should occur for any QPC orientation, fabricated on any crystal plane, including the high symmetry (100) plane \cite{NichelePRL14}.

In conclusion, Zeeman splitting measurements of 1D subbands were carried out for three hole QPCs on the high symmetry (100) plane and the low symmetry (311) plane of GaAs.  For the QPCs on the (311) plane, strong anti-crossing of 1D subbands was observed in an in-plane field.  We presented a theoretical framework showing that these anti-crossings occur due to the interplay between the SOI terms (Rashba and Dresselhaus) and the off-diagonal hole $g$ tensor.  Our experiment demonstrates independent detection and control of the Dresselhaus SOI, and provides new insights into spin-orbit effects in quantum confined hole systems.

\begin{acknowledgments}
The authors acknowledge the late J. Cochrane for technical support, and thank T. Li for useful discussions.  YH acknowledges support by KAKENHI grant (Nos. 26287059 and 15H05867) and CSRN Tohoku University. DR and ADW acknowledge support from BMBF - Q.com-H  16KIS0109.  This work was supported by the Australian Research Council under the DP scheme, and was performed in part using facilities of the NSW Node of the Australian National Fabrication Facility.
\end{acknowledgments}


\begin{thebibliography}:

\bibitem{DattaApl90} S. Datta and B. Das, \textit{Appl. Phys. Lett.} {\bf 56}, 665 (1990).

\bibitem{LossPRA98} D. Loss and D. P. DiVincenzo, \textit{Phys. Rev. A.} \textbf{57}, 120 (1998).

\bibitem{WolfSci01} S. A. Wolf, D. D. Awschalom, R. A. Buhrman, J. M. Daughton, S. von Molnar, M. L. Roukes, A. Y. Chtchelkanova, and D. M. Treger, \textit{Science} {\bf 294}, 1488 (2001).

\bibitem{Awsbook02} D. D. Awschalom, N. Samarth, D. Loss, Eds., \textit{Semiconductor Spintronics and Quantum Computation} (Springer-Verlag, Berlin, Germany, 2002).

\bibitem{WinklerBook03} R. Winkler, \textit{Spin-orbit coupling effects in two-dimensional electron and hole systems}, (Springer Tracts in Modern Physics, Vol. 191, Springer, Berlin, 2003).

\bibitem{Dress} G. Dresselhaus, \textit{Phys. Rev.} \textbf{100}, 580 (1955).

\bibitem{Rashba} Y. A. Bychkov and E. I. Rashba, \textit{J. Phys. C} \textbf{17}, 6039 (1984).

\bibitem{WieckPRL} A. D. Wieck, E. Batke, D. Heitmann, J. P. Kotthaus and E. Bangert, \textit{Phys. Rev. Lett.} {\bf 53}, 493 (1984).

\bibitem{DlossPRL07} D. V. Bulaev and D. Loss, \textit{Phys. Rev. Lett.} {\bf 98}, 097202 (2007).


\bibitem{GoloPRB06} V. N. Golovach, M. Borhani, and D. Loss, \textit{Phys Rev. B}, \textbf{74}, 165319 (2006).






\bibitem{Helix} J. D. Koralek, C. P. Weber, J. Orenstein, B. A. Bernevig, Shou-Cheng Zhang, S. Mack and D. D. Awschalom, \textit{Nature} {\bf 458}, 610 (2009).

\bibitem{111} L. Wang and M. W. Wu, \textit{Phys. Rev. B} {\bf 85}, 235308 (2012).

\bibitem{Sasaki} A. Sasaki,	S. Nonaka, Y. Kunihashi, M. Kohda, T. Bauernfeind, T. Dollinger, K. Richter and J. Nitta, \textit{Nat. Nano.} {\bf 9}, 703–709 (2014).

\bibitem{DlossPRB05} D. V. Bulaev and D. Loss, \textit{Phys. Rev. Lett.} {\bf 95}, 076805 (2005).

\bibitem{AwongPRB10} A. Wong and F. Mireles, \textit{Phys. Rev. B} {\bf 81}, 085304 (2010).


\bibitem{ClarkeJAP06} W. R. Clarke, A. P. Micolich, A. R. Hamilton, M. Y. Simmons, K. Muraki and Y. Hirayama, \textit{J. Appl. Phys.} {\bf 99}, 023707 (2006).

\bibitem{ChenNJP10} J. C. H Chen, O. Klochan, A. P. Micolich, A. R. Hamilton, T. P. Martin, L. H. Ho, U. Z\"{u}licke, D. Reuter and A. D. Wieck, \textit{New. J. Phys.} {\bf 12}, 033043 (2010).

\bibitem{MicoJPCM11} A. P. Micolich, \textit{J. Phys. Cond. Mat.} {\bf 23}, 443201 (2011).

\bibitem{OlgaPRL14} O. Goulko, F. Bauer, J. Heyder, and J. von Delft, \textit{Phys. Rev. Lett.} \textbf{113}, 266402 (2014).

\bibitem{IqbalNat13} M. J. Iqbal, R. Levy, E. J. Koop, J. B. Dekker, J. P. de Jong, J. H. M. van der Velde, D. Reuter, A. D. Wieck, R. Aguado, Y. Meir and C. H. van der Wal, \textit{Nature} \textbf{501}, 79-83 (2013).

\bibitem{NichelePRL14} F. Nichele, S. Chesi, S. Hennel, A. Wittmann, C. Gerl, W. Wegscheider, D. Loss, T. Ihn and K. Ensslin, \textit{Phys. Rev. Lett.} \textbf{113}, 046801 (2014).

\bibitem{AshwinNL13} A. Srinivasan, L. A. Yeoh, O. Klochan, T. P. Martin, J. C. H. Chen, A. P. Micolich, A. R. Hamilton, D. Reuter and A. D. Wieck, \textit{Nano Lett.} \textbf{13}, 148-152 (2013).

\bibitem{KomijaniEPL13} Y. Komijani, M. Csontos, I. Shorubalko, U. Zulicke, T. Ihn, K. Ensslin, D. Reuter and A. D. Wieck \textit{Eur. Phys. Lett.} {\bf 102}, 37002 (2013).

\bibitem{suppmat} See supplemental material.

\bibitem{Liz} E. Marcellina, A. R. Hamilton, R. Winkler and D. Culcer, \textit{Phys. Rev. B.} {\bf 95}, 075305 (2017).

\bibitem{WinklerSST08} R. Winkler, D. Culcer, S. J. Papadakis, B. Habib and M. Shayegan, \textit{Semicond. Sci. Technol. } {\bf 23}, 114017 (2008).



\bibitem{AshwinPRB16} A. Srinivasan, K. L. Hudson, D. Miserev, L. A. Yeoh, O. Klochan, K. Muraki, Y. Hirayama, O. P. Sushkov, and A. R. Hamilton, \textit{Phys. Rev. B} \textbf{94}, 041406(R) (2016).


\bibitem{YeohPRL14} L. A. Yeoh, A. Srinivasan, O. Klochan, R. Winkler, U. Z\"{u}licke, M. Y. Simmons, D. A. Ritchie, M. Pepper and A. R. Hamilton, \textit{Phys. Rev. Lett.} {\bf 113}, 236401 (2014).



\end{thebibliography}
\end{document}